             \documentstyle[12pt,epsf]{article}
\begin{document}\begin{flushright}\thispagestyle{empty}
%OUT--4102--8?\\
MZ--TH/00--28\\ hep-th/0012146\\  Nov 20 2000
\end{flushright}\vspace*{2mm}\begin{center}{
                                                    \Large\bf
Exact solutions of Dyson-Schwinger equations for iterated one-loop
integrals and propagator-coupling duality
\\[3pt]
                                                    }\vglue 10mm{\large\bf
D.~J.~Broadhurst$^{1)}$ and D.~Kreimer$^{2)}$       }\end{center}\vfill
                                                  \noindent{\bf Abstract}\quad
The Hopf algebra of undecorated rooted trees has tamed the
combinatorics of perturbative contributions, to anomalous
dimensions in Yukawa theory and scalar $\phi^3$ theory, from all
nestings and chainings of a primitive self-energy subdivergence.
Here we formulate the nonperturbative problems which these
resummations approximate. For Yukawa theory, at spacetime
dimension $d=4$, we obtain an integrodifferential Dyson-Schwinger
equation and solve it parametrically in terms of the complementary
error function. For the scalar theory, at $d=6$, the
nonperturbative problem is more severe; we transform it to a
nonlinear fourth-order differential equation. After intensive use
of symbolic computation we find an algorithm that extends both
perturbation series to 500 loops in 7 minutes.
%We estimate the corresponding critical scalar coupling.
Finally, we establish the propagator-coupling duality underlying
these achievements  making use of the Hopf structure of Feynman
diagrams. \vfill\footnoterule\noindent {\footnotesize $^1$)
D.Broadhurst@open.ac.uk;
http://physics.open.ac.uk/$\;\widetilde{}$dbroadhu\\ Physics Dept,
Open University, Milton Keynes MK7 6AA, UK\\ $^2$)
Dirk.Kreimer@uni-mainz.de;
http://dipmza.physik.uni-mainz.de/$\;\widetilde{}$kreimer/homepage.html\\
Heisenberg Fellow, Physics Dept, Univ.\ Mainz, 55099 Mainz,
Germany}
\newpage\setcounter{page}{1}

\section{Introduction}

In~\cite{exp,BK} we developed the Hopf-algebra techniques
of~\cite{DK,CK,Over,Chen} to tame the combinatoric explosion of
perturbation theory. For Yukawa theory at spacetime dimension
$d=4$, and also for a scalar $\phi^3$ theory at its critical
dimension $d=6$, we resummed all nestings and chainings of a
divergent self-energy skeleton, using the Hopf algebra of
undecorated rooted trees, to progress beyond the
rainbow~\cite{Del4,Del6} and chain~\cite{Chain,BG} approximations
for anomalous dimensions. Pad\'e-Borel resummation to 30 loops
gave stable results at a Yukawa coupling $g=30$, enabling
comparison with resummation methods. However, our previous work
left unanswered two pertinent questions.
\begin{enumerate}
\item What are the nonperturbative problems whose
perturbations~\cite{exp,BK} were developed?
\item Are there limits on the coupling strength for which
resummation may be performed?
\end{enumerate}
Here we answer both questions. Our results provide a stringent testing ground
 \cite{Jen} for confronting resummations of symbolically computed perturbation
theory with high-precision numerical analyses of the corresponding
nonperturbative integrodifferential equations.

Sect.~2 derives the nonperturbative problems
as integrodifferential Dyson-Schwinger equations.
These may be transformed to nonlinear differential equations,
of second order in the Yukawa case and fourth order in the scalar
case. The anomalous dimension is then defined, nonperturbatively,
by the unique value of the derivative of the renormalized self-energy
at the subtraction point $q^2=\mu^2$ that guarantees a well-defined
self-energy at all euclidean momenta with $0<q^2<\mu^2$. For the Yukawa case a
parametric solution is obtained in terms of the complementary error function,
erfc, with the anomalous dimension $\widetilde\gamma$ at coupling
$a=(g/4\pi)^2$ given by the implicit condition
\begin{equation}
\sqrt{a\over2\pi}=\exp(p_0^2)\,{\rm erfc}(p_0)\,;
\quad p_0:={\widetilde\gamma+2\over\sqrt{2a}}
\label{Yuksol}
\end{equation}
which we solve by a Newton-Raphson method. The asymptotic
perturbation series
\begin{equation}
\widetilde\gamma\simeq\sum_{n>0}
\widetilde{G}_n{(-a)^n\over2^{2n-1}}
\label{Yukasy}
\end{equation}
has a coefficient at $n+1$ loops given recursively by
\begin{equation}
\widetilde{G}_{n+1}=\delta_{n,0}+n\sum_{k=1}^n
\widetilde{G}_{k}\widetilde{G}_{n+1-k}\,.
\label{Yukrec}
\end{equation}
The corresponding problem in the scalar theory entails boundary conditions
on a nonlinear fourth-order differential equation. We show how to
develop the perturbation series
\begin{equation}
\gamma\simeq\sum_{n>0}
G_n{(-a)^n\over6^{2n-1}}
\label{phiasy}
\end{equation}
for the scalar anomalous dimension $\gamma$, far
beyond the 30 loops obtained in~\cite{exp}. After attempting a variety of
methods, we found one that delivers 500 loops in 7 minutes,
giving the 1675-digit integer $G_{500}$ of Table~1.

In the Yukawa case, an obstacle exists at the critical coupling
constant $g_{\rm crit}:=2(2\pi)^{3/2}\approx31.5$, corresponding
to a critical anomalous dimension $\widetilde\gamma_{\rm
crit}:=-2$. The asymptotic series
\begin{equation}
\widetilde\gamma\simeq
2\sum_{n>0}(2n-1)!!\left({-a\over(\widetilde\gamma+2)^2}\right)^n
\label{erfasy}
\end{equation}
is obtained as $a\to0$. It is clearly vacuous as $\widetilde\gamma$
approaches $-2$. At such strong couplings, one should use our new
nonperturbative result~(\ref{Yuksol}), with
\begin{equation}
{\rm erfc}(p_0):=
{2\over\sqrt\pi}\int_{p_0}^\infty dp\,\exp(-p^2)
=1-{2p_0\over\sqrt\pi}\sum_{n=0}^\infty{(-p_0^2)^n\over n!(2n+1)}\,.
\label{erfc}
\end{equation}
Having thus solved the Yukawa problem at all couplings,
we use its perturbative formulation to motivate methods
for the more severe fourth-order scalar problem, whose Dyson-Schwinger
equation indicates an obstacle at $\gamma=-1$.

\section{Dyson-Schwinger analyses}

In Yukawa theory, with an interaction term
$g\overline\psi\sigma\psi$ at $d=4$, we consider the dimensionless
renormalized self-energy term $\widetilde\Sigma(q^2)$ in the
inverse propagator $q\llap{/\kern-0.5pt}(1-\widetilde\Sigma(q^2))$
of a massless fermion field $\psi$ at euclidean 4-momentum $q$.
The subtraction is performed
at $q^2=\mu^2$, with $\widetilde\Sigma(\mu^2)=0$.
Then $\widetilde\gamma$
is the value of $d\log(1-\widetilde\Sigma(q^2))/d\log(q^2)$ at $q^2=\mu^2$.
In the scalar theory at $d=6$ we consider an interaction
$g\phi^\dagger\sigma\phi$ of the massless neutral scalar field $\sigma$
with a massless charged scalar field $\phi$, whose inverse propagator is
$q^2(1-\Sigma(q^2))$.

The infinite set of Feynman diagrams which we resum
comprises all those whose subdivergences result from nestings and
chainings of the one-loop self-energy skeleton. A 12-loop example is given
in Fig.~1, with a subdivergence structure encoded by the
undecorated rooted tree of Fig.~2. The Hopf algebraic methods
of~\cite{BK} gave the integer coefficients $\widetilde{G}_n$
and $G_n$ of the perturbation series~(\ref{Yukasy},\ref{phiasy})
up to $n=12$ loops, with a coupling $a:=g^2/(4\pi)^{d/2}$. In~\cite{exp}
we were able to extend this analysis to 30 loops, obtaining
\begin{eqnarray}
\widetilde{G}_{30}&=&
10272611586206353744425870217572111879288
\label{Yuk30}\\
G_{30}&=&
19876558632009586773182109989526780486481329823560105761256963720
\quad{}\label{phi30}
\end{eqnarray}
each of which subsumes $4.6\times10^{20}$ subtractions in the BPHZ
formalism~\cite{BPHZ}. At that point, our dimensionally
regularized method hit a limited imposed by memory requirements,
since we had to store a triple series in powers of the coupling
$a$, the logarithm $\log(q^2/\mu^2)$, and a dimensional
regularization parameter $\varepsilon$.
\begin{figure}\centering\fbox{
\epsfysize=3cm \epsfbox{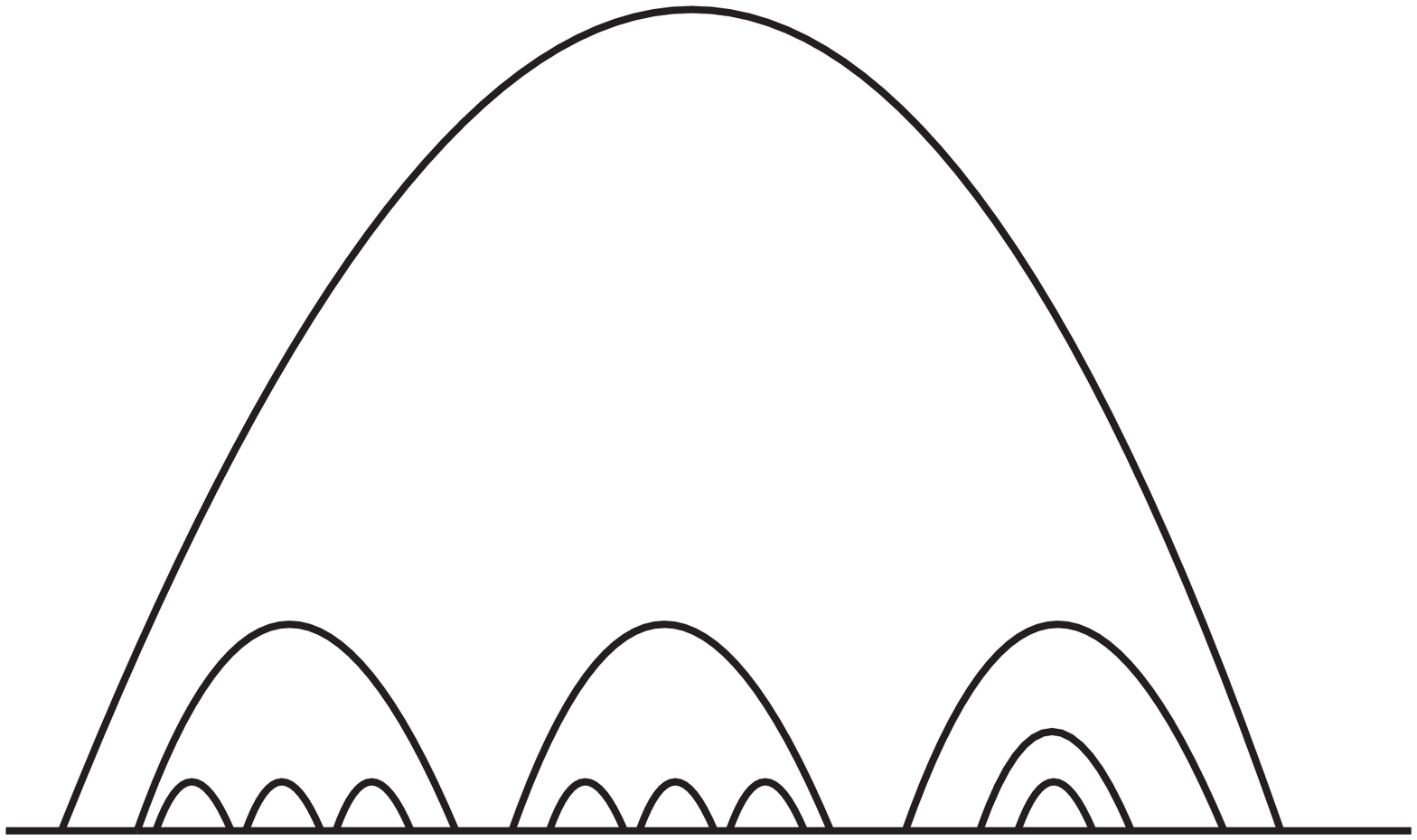}} \caption{A twelve-loop example
of the type of diagrams considered.}
\end{figure}
\begin{figure}
\centering\fbox{ \epsfysize=3cm \epsfbox{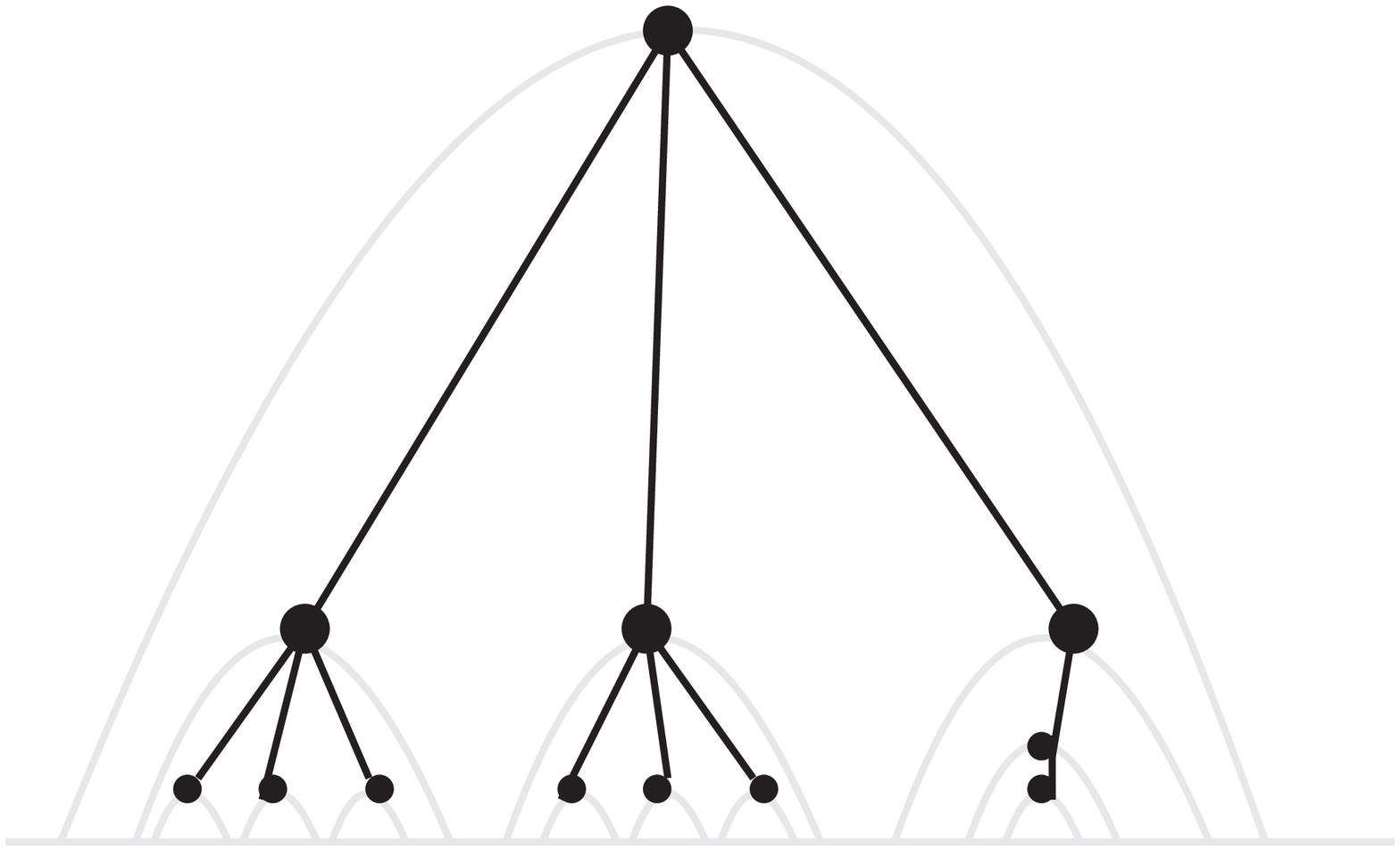}} \caption{The
corresponding tree structure which determines the necessary
subtractions to get local counterterms.}
\end{figure}

We shall greatly improve on the previous analyses~\cite{exp,BK},
by using Dyson-Schwinger methods. First we show how to reduce the
triple series to a double series, by dispensing with dimensional
regularization. Then we describe the symbolic computation proof
methods that led to discovery of a single-series method, which
dispenses with the logarithm. They culminate in a
propagator-coupling duality which we then prove thanks to the Hopf
algebra structure.

\subsection{Integrodifferential equations}

Modulo divergences,
the Yukawa problem corresponds to the Dyson-Schwinger equation
\begin{equation}
q^2\widetilde\Sigma(q^2)={a\over\pi^2}
\int{d^4l\over l^2-l^2\widetilde\Sigma(l^2)}
{q\cdot l\over(q+l)^2}-\mbox{subtractions}
\label{Yukint}
\end{equation}
with subtractions that give $\widetilde\Sigma(\mu^2)=0$.

%xxxx
Let us pause for a moment and consider the structure of this
equation. Its design guarantees that it represents a sum over all
undecorated rooted trees: If we let $$ q^2\widetilde{\bf
\Sigma}(q^2)[t]=\frac{a}{\pi^2}\int d^4l \widetilde{\bf
\Sigma}(l^2)[B_-(t)]\frac{q\cdot l}{l^2(q+l)^2} $$ be, for each
$q^2$, a character on the Hopf algebra of undecorated rooted trees
(which implies to define $\widetilde{\bf \Sigma}(q^2)[e]:=1$ and
$\widetilde{\bf \Sigma}(q^2)[t_1 t_2]:=\widetilde{\bf
\Sigma}(q^2)[t_1]\widetilde{\bf\Sigma}(q^2)[t_2]$), then  we can
regard $$\widetilde{B_+}:=\frac{a}{\pi^2}\int d^4 l \frac{q\cdot
l}{l^2(q+l)^2}$$ as the corresponding representation of the $B_+$
operator with coproduct $\Delta(T)=T\otimes 1+({\rm id}\otimes
B_+)\Delta(B_-(T))$ \cite{BK},  $$ \widetilde{B_+}(\widetilde{\bf
\Sigma})(q^2)[t]:=\widetilde{\bf\Sigma}(q^2)[B_+(t)]=\frac{a}{\pi^2}\int
d^4l \widetilde{\bf\Sigma}(l^2)[t]\frac{q\cdot l}{l^2(q+l)^2}. $$

The Dyson--Schwinger equation can then be written as $$
q^2\widetilde{\bf \Sigma}(q^2)[X]=\frac{a}{\pi^2} \int d^4 l
\widetilde{\bf \Sigma}(l^2)[1/(1-X)]\frac{q\cdot l}{l^2(q+l)^2},$$
where the series in Hopf algebra elements $X$ is defined by
$X=aB_+(1/(1-X))$, a form which we will use later on.

As we are confronting a character of the Hopf algebra of
undecorated rooted trees, prominent in the study of differential
equations in the context of Runge--Kutta methods, we expect to be
able to find a solution to the Dyson--Schwinger equation turning
it into a differential equation. This is indeed the case. We
proceed by first turning the problem into a integrodifferential
equation.
%xxxx

The integration over the 4-dimensional euclidean loop momentum, $l$,
factors into an angular and a radial part. The angular integration of
\begin{equation}
{q\cdot l\over(q+l)^2}={1\over2}-{q^2+l^2\over2(q+l)^2}
\label{sp}
\end{equation}
is performed by using the 4-dimensional angular average
\begin{equation}
q^2l^2\left\langle{1\over(q+l)^2}\right\rangle_{d=4}
=\min(q^2,l^2)\,.
\label{av4}
\end{equation}
We split the radial
integral over $y:=l^2$ into parts with $y<q^2$ and $y>q^2$. For the latter
we fix the subtractions by requiring that $\widetilde\Sigma(\mu^2)=0$,
obtaining the Dyson-Schwinger equation
\begin{eqnarray}
\widetilde\Sigma(q^2)&=&{a\over2}\int_{\mu^2}^{q^2}
{dy\over y-y\widetilde\Sigma(y)}+\widetilde{F}(\mu^2)-\widetilde{F}(q^2)
\label{YukDS}\\
\widetilde{F}(x)&:=&
{a\over2}\int_0^x{dy\over y-y\widetilde\Sigma(y)}\left(y\over x\right)^2\,.
\label{YukF}
\end{eqnarray}
Differentiating w.r.t. $x:=q^2$ we obtain the integrodifferential
equation
\begin{equation}
x^3{d\over dx}\widetilde\Sigma(x)=a
\int_0^x{y\,dy\over1-\widetilde\Sigma(y)}
\label{Yukid}
\end{equation}
which makes no reference to the subtraction point.

To repeat this analysis in the scalar case, we use the
6-dimensional angular average
\begin{equation}
q^2l^2\left\langle{1\over(q+l)^2}\right\rangle_{d=6}
=N-{N^3\over3q^2l^2}\,;\quad N:=\min(q^2,l^2)
\label{av6}
\end{equation}
which gives the Dyson-Schwinger equation
\begin{eqnarray}
\Sigma(q^2)&=&{a\over6}\int_{\mu^2}^{q^2}
{dy\over y-y\Sigma(y)}+F(\mu^2)-F(q^2)
\label{phiDS}\\
F(x)&:=&
{a\over6}\int_0^x{dy\over y-y\Sigma(y)}\left({y^3\over x^3}
-3{y^2\over x^2}+3{y\over x}\right)
\label{phiF}
\end{eqnarray}
and hence the integrodifferential equation
\begin{equation}
x^4{d\over dx}\Sigma(x)={a\over2}
\int_0^x{(x-y)^2dy\over1-\Sigma(y)}
\label{phiid}
\end{equation}
which likewise makes no reference to the subtraction point.
We require the solution to~(\ref{phiid}) that is regular for $0<x\le\mu^2$,
with $\Sigma(\mu^2)=0$. Then the scalar anomalous dimension is
\begin{equation}
\gamma:=
\left.{d\log(1-\Sigma(q^2))\over d\log(q^2)}\right|_{q^2=\mu^2}=
-{a\over2}\int_0^1{(1-r)^2dr\over1-\Sigma(\mu^2r)}\,.
\label{def}
\end{equation}

\subsection{Multiloop integers from nonlinear differential equations}

By differentiating~(\ref{Yukid}) once and~(\ref{phiid}) three times we
obtain
\begin{equation}
(1-\widetilde\Sigma(q^2))\widetilde{P}(D)\widetilde\Sigma(q^2)
=a=(1-\Sigma(q^2))P(D)\Sigma(q^2)
\label{ode}
\end{equation}
with polynomials of the differential operator $D:=d/d\log(q^2)$ given by
\begin{eqnarray}
\widetilde{P}(D)&:=&D(D+2)
\label{YukD}\\
\quad P(D)&:=&D(D+1)(D+2)(D+3)
\label{phiD}
\end{eqnarray}
and already encountered in analyses of rainbow and chain
contributions. The rainbow diagrams generate
contributions to the anomalous dimensions that solve
\begin{equation}
\widetilde{P}(-\widetilde\gamma_{\rm rainbow})
=a=P(-\gamma_{\rm rainbow})
\label{rain}
\end{equation}
namely~\cite{Del4,Del6}
\begin{eqnarray}
\widetilde\gamma_{\rm rainbow}=1-\sqrt{1+a}
&=&-\frac{a}{2}+\frac{a^2}{2^3}
-2\frac{a^3}{2^5}+5\frac{a^4}{2^7}+\cdots
\label{R4}\\
\gamma_{\rm rainbow}=\frac{3-\sqrt{5+4\sqrt{1+a}}}{2}
&=&-\frac{a}{6}+11\frac{a^2}{6^3}-206\frac{a^3}{6^5}+\cdots
\label{R6}
\end{eqnarray}
whereas insertion of chains in the one-loop diagram gives~\cite{exp}
\begin{eqnarray}
\widetilde\gamma_{\rm chain}&=&-2\int_0^\infty\exp(-2x/a){x\,dx
\over\widetilde{P}(x)}
\simeq-\frac{a}{2}+\frac{a^2}{2^3}
-2\frac{a^3}{2^5}+6\frac{a^4}{2^7}+\cdots
\label{C4}\\
\gamma_{\rm chain}&=&-6\int_0^\infty\exp(-6x/a){x\,dx
\over P(x)}
\simeq-\frac{a}{6}+11\frac{a^2}{6^3}-170\frac{a^3}{6^5}+\cdots
\label{C6}
\end{eqnarray}
At $n<3$ loops, rainbow and chain contributions correspond to
same rooted trees. At $n=3$ loops, there are 2 rooted trees and the
full Hopf algebra gives the sum of the rainbow and chain contributions,
with $\widetilde{G}_3=2+2=4$
and $G_3=206+170=376$. At $n>3$ loops, there are essentially new terms
in the Hopf algebra. For example, $\widetilde{G}_4=27>5+6$ receives
contributions from all 4 rooted trees with 4 nodes.

Thanks to the differential equations in~(\ref{ode}) we can achieve a
dramatic speedup of the dimensionally regularized methods in~\cite{exp,BK}.
The $n$-loop
contribution to $\Sigma(q^2)$ has the form
$a^n\sum_{k=1}^n C(n,k)L^k$, with $L:=\log(q^2/\mu^2)$. Substituting this
Ansatz into the nonlinear fourth-order differential equation
$a=(1-\Sigma(q^2))P(D)\Sigma(q^2)$,
we require that all powers of the log disappear
on the r.h.s.\ and that only the first power of the coupling survives.
Then the integer $G_n=6(-36)^{n-1}C(n,1)$ is obtained by iterative
solution of a system of equations that gives $n$ linear combinations
of $\{C(n,k)\mid 1\le k\le n\}$ in terms of $O(n^3)$ products of
$\{C(m,k)\mid 1\le k\le m<n\}$.

This Dyson-Schwinger algorithm may be implemented by the {\sc reduce} procedure
\begin{verbatim}
procedure HopfScalar(n); for m:=1:n do
<<s m:=for k:=1:m sum c k*log x^k;
d m:=if m=1 then 1 else for k:=1:m-1 sum d k*s(m-k);
sol:=first solve coeff(x*df(x^3*s m,x,4)-d m,log x);
s m:=if m=1 then log x/6 else (s m where sol);
write G m:=6(-36)^(m-1)*sub(x=1,df(s m,x))>>;
\end{verbatim}
which invokes the procedures {\tt solve}, {\tt coeff} and {\tt df},
to solve for the coefficients of the results of differentiations.
It yields the 30-loop coefficient~(\ref{phi30}) in 6 seconds, running version
3.7~\cite{Red} of {\sc reduce} on a 500 MHz alpha machine.
By way of comparison, the dimensionally regularized 30-loop calculation
in~\cite{exp} took 8 hours, on the same machine.
The origin of the speedup is clear:
here we have only a double expansion, in the coupling and
the logarithm, whereas in~\cite{exp,BK}
we also had a Laurent series in $\varepsilon$, where the limit
$\varepsilon\to0$ was admissible only at the final stage of the
dimensionally regularized BPHZ procedure.
Now that we have a nonperturbative Dyson-Schwinger formulation,
we can dispense with the BPHZ complexities.

Note that, even using the new method,
the time to reach $n$ loops grows rather rapidly
with $n$. At each iteration with $m\le n$ we solve a
system of $m$ linear equations and hence perform
a number of elementary operations that grows like $m^3$.
The main cost comes from $O(m^3)$ multiplication of integers
with $O(m\log m)$ digits at the $m$-th step of the iteration.
It follows
that the time to reach $n$ loops scales like $n^c$, modulo
logarithms of $n$, with an exponent $c\ge 5$, in any implementation,
and with $c\le6$, in any sensible implementation. To achieve the lower
bound $c=5$ one might need to use fast Fourier transforms.
We devised {\sc reduce} code, more structured than that above,
whose timing scaled, empirically, with an exponent $c\approx 5.3$,
for $n=O(100)$. After a couple of days we reached 420 loops
and checked the integer $G_{420}$ by independent
code that used David Bailey's {\sc mpfun}
package~\cite{DHB}. As might have been expected, Bailey's
multiple-precision extension of {\sc fortran}
was somewhat faster.
Table~1 gives the result for $G_{500}$, obtained by running
{\sc mpfun} for 21.5 hours.

We now chart the discovery of a second dramatic speedup, which
reduces the number of multiplications of large integers at the
$m$-th step from $O(m^3)$ to merely $O(m)$.

\subsection{Parametric solution of the Yukawa problem}

For the Yukawa problem we work with the variable $z:=(q^2/\mu^2)^2$ and define
\begin{equation}
\widetilde{G}(z):=\sqrt{2/a}
\left(z-z\widetilde\Sigma\left(\mu^2z^{1/2}\right)\right)
\label{YukG}
\end{equation}
normalized so that $\widetilde{G}(1)=\sqrt{2/a}$.
Then the differential equation is simply
\begin{equation}
2\widetilde{G}\widetilde{G}^{\prime\prime}=-1
\label{Yukb}
\end{equation}
where primes denote differentiation w.r.t.\ $z$. This is easily
integrated, to give
\begin{equation}
(\widetilde{G}^\prime)^2=-\log\widetilde{G}+\mbox{constant}\,.
\label{Yukif}
\end{equation}
The next step is to work with the parameter $p:=\widetilde{G}^\prime$,
in terms of which
\begin{equation}
\widetilde{G}=\sqrt{2/a}\exp(p_0^2-p^2)
\label{YukGp}
\end{equation}
where $p_0$ is the value of $p$ for which $z=1$ and hence $q^2=\mu^2$.
Then a parametric solution is obtained by determining
\begin{equation}
\widetilde\alpha(p):=z/\widetilde{G}
\label{Yukzp}
\end{equation}
which satisfies the first-order equation
\begin{equation}
\widetilde\alpha={1\over p}+{1\over2p}{d\widetilde\alpha\over dp}
\label{Yuka}
\end{equation}
whose iterative solution gives the asymptotic series
\begin{equation}
\widetilde\alpha(p)\simeq{1\over p}+{1\over p}
\sum_{n>0}{(2n-1)!!\over(-2p^2)^n}\,.
\label{Yukap}
\end{equation}
From~(\ref{Yukzp}) and the derivative of~(\ref{YukG}) at $z=1$, we determine
\begin{equation}
\widetilde\alpha(p_0)=\sqrt{a\over2}\,;\quad
p_0:={\widetilde{\gamma}+2\over\sqrt{2a}}
\label{Yukp0}
\end{equation}
and hence obtain the asymptotic series~(\ref{erfasy}) at $p=p_0$.

As advertised in the introduction,
we can now give the complete nonperturbative solution of the $d=4$ case,
by noting that the asymptotic series~(\ref{Yukap}) is that of
\begin{equation}
\widetilde\alpha(p)=\sqrt\pi\exp(p^2)\,{\rm erfc}(p)
=2\int_p^\infty ds\exp(p^2-s^2)\,.
\label{Yukapis}
\end{equation}
Tidying up, we then obtain the parametric solution
\begin{eqnarray}
\widetilde\Sigma(q^2)&=&1-{\sqrt{a/2\pi}\over\exp(p^2)\,{\rm erfc}(p)}
\label{p1}\\
q^2&=&\mu^2\left({{\rm erfc}(p)
\over{\rm erfc}(p_0)}\right)^{1/2}
\label{p2}
\end{eqnarray}
with the condition~(\ref{Yuksol})
determining the anomalous dimension, even in the strong-coupling regime
$g>g_{\rm crit}:=2(2\pi)^{3/2}\approx31.5$ where
$a>2\pi$ and hence $p_0:=(\widetilde\gamma+2)/\sqrt{2a}<0$.

Finally we obtain the elegant recursion~(\ref{Yukrec})
for the coefficients of the perturbation series~(\ref{Yukasy}),
by turning~(\ref{Yuka}), at $p=p_0$, into a nonlinear differential equation
for the anomalous dimension. The result is conveniently written as
\begin{equation}
2\widetilde\gamma=-a+a^2{d\over d a}{\widetilde\gamma^2\over a}
\label{conv}
\end{equation}
which proves~(\ref{Yukrec}) and generates the 500-loop Yukawa
coefficient in 10 seconds. We remind the reader that this sums all
diagrams at the 500-loop level generated by our Dyson--Schwinger
equations, of which there are as many as there are rooted trees
with 500 vertices. Finally, we note that at g=30 the
nonperturbative result for $\tilde\gamma$, to 120 digits,
is\\[3mm] {\tt
-1.8520278058795936576120591680015771764936420535922291037938512550071476\\
394977783846264445624911569322962723432940741627702,\\[3mm]}
obtained by a Newton-Raphson method and in excellent agreement
with resummation techniques \cite{Jen}.

\subsection{Asymptotic parametric solution of the scalar problem}

In the scalar case, at $d=6$, we work with the variable $y:=q^2/\mu^2$
and define
\begin{equation}
G(y):=\sqrt{6/a}\left(y-y\Sigma\left(\mu^2y\right)\right)
\label{G}
\end{equation}
which transforms the fourth-order equation for $\Sigma$ to
\begin{equation}
GG^{\prime\prime}+{G\over6}{d\over dy}\left(5+y{d\over dy}\right)
y G^{\prime\prime}=-1
\label{4o}
\end{equation}
where primes denote differentiation w.r.t.\ $y$.
Taking $p:=G^\prime$ as the parameter, we define
\begin{equation}
\alpha(p):=y/G\,;\quad \beta(p):=-GG^{\prime\prime}\label{ab}
\end{equation}
with the formal solution
\begin{equation}
\alpha(p)=\int_p^\infty{ds\over\beta(s)}
\exp\left(-\int_p^s{t\,dt\over\beta(t)}\right)
\label{formal}
\end{equation}
generalizing~(\ref{Yukapis}), which was the Yukawa case
$\widetilde\beta=1/2$. The
differential equations
\begin{eqnarray}
\alpha&=&{1\over p}+{\beta\over p}{d\alpha\over dp}
\label{1o}\\
\beta&=&1+{\beta\over6}{d\over dp}\left(5-\alpha\beta{d\over dp}\right)
\alpha\beta
\label{2o}
\end{eqnarray}
may be solved iteratively, to yield the asymptotic series
\begin{eqnarray}
\alpha(p)&\simeq&{1\over p}+{6\over p}\sum_{n>0}{A_n\over(-6p^2)^n}
\label{ait}\\
\beta(p)&\simeq&1+\sum_{n>0}{B_n\over(-6p^2)^n}
\label{bit}
\end{eqnarray}
where $A_n$ and $B_n$ are positive integers, with $A_1=1$ and $B_1=5$,
obtained by inspection, leading to $A_2=23$ and $B_2=4\times43$, and so on.
Diligent symbolic programming reduces the main computational
burden to $O(n)$ multiplications of large integers, at the $n$-th
iteration of~(\ref{1o},\ref{2o}).

By this means we developed 500 terms of the asymptotic series
\begin{equation}
\gamma\simeq6\sum_{n>0}A_n\left({-a\over(6\gamma+6)^2}\right)^n
\label{geasy}
\end{equation}
in a minute, while a day was needed to compute $G_{500}$,
by the double-series method.

\subsection{Differential equation for the scalar anomalous dimension}

In the Yukawa case, we succeeded in computing $\widetilde{G}_n$
via $O(n)$ large-integer multiplications in~(\ref{Yukrec}). Similarly,
in the scalar case, we programmed the iterations of~(\ref{1o},\ref{2o})
so as to obtain $A_n$ via $O(n)$ multiplications at the $n$-th step.

Yet there is a glaring discrepancy between the Yukawa and scalar
analyses, thus far. The procedure for obtaining $G_n$ from $A_n$
appears to involve $O(n^3)$ multiplications. Thus we have not yet
improved on the double-series method, in the scalar case. We
eventually remedied this problem, by a sequence of
computer-algebra explorations and finally reduced it to the
procedure {\tt DysonScalar} to which we now turn.

First, we sought generalizations of the method of~(\ref{1o},\ref{2o}).
The factors of~(\ref{phiD}) reveal that so far we have
exploited only one of three possibilities. More generally,
we obtained
\begin{eqnarray}
\alpha_k&=&{1\over p}+{\beta_k\over p}{d\alpha_k\over dp}
\simeq{1\over p}+{6\over p}\sum_{n>0}{A_{n,k}\over(-6p^2)^n}
\label{1ok}\\
\beta_k&=&1+{\beta_k\over6}{d\over dp}\left(11k-6
-k^3\alpha_k\beta_k{d\over dp}\right)
\alpha_k\beta_k
\label{2ok}\\
\gamma&\simeq&6k\sum_{n>0}A_{n,k}\left({-k a\over(6\gamma+6k)^2}\right)^n
\label{gk}
\end{eqnarray}
which is valid for $k=1,2,3$, but for no other value. The
analysis in~(\ref{1o},\ref{2o},\ref{geasy}) corresponds
to the case with $k=1$.
%The cases with $k=2$ and $k=3$ will likewise
%be used in the numerical analysis of sect.~3.

To check that~(\ref{2ok}) carries the same information for $k=1,2,3$,
we used computer algebra
to transform it to a nonlinear third-order differential equation
for $\gamma$, invoking the constraint $(k-1)(k-2)(k-3)=0$
to obtain
\begin{eqnarray}
&&8a^3\gamma\left\{\gamma^2\gamma^{\prime\prime\prime}
+4\gamma\gamma^\prime\gamma^{\prime\prime}
+(\gamma^\prime)^3\right\}
+4a^2\gamma\left\{2\gamma(\gamma-3)\gamma^{\prime\prime}
+(\gamma-6)(\gamma^\prime)^2\right\}
\nonumber\\&&{}
+2a\gamma(2\gamma^2+6\gamma+11)\gamma^\prime
-\gamma(\gamma+1)(\gamma+2)(\gamma+3)=a
\label{3o}
\end{eqnarray}
where the primes denote differentiation w.r.t.\ the coupling $a$.
This seemingly baroque equation then remedies the previous discrepancy between
the scalar and Yukawa cases, allowing an iterative determination of
the coefficients $G_n$ via $O(m)$ large-integer multiplications
at each $m\le n$. The iterative procedure is based on the rearrangement
\begin{eqnarray}
6\gamma&=&-a+11a^2{d\over d a}{\gamma^2\over a}-6T_3+T_4
\label{T34}\\
T_3&:=&4a^2\gamma\left\{\gamma\gamma^{\prime\prime}
+(\gamma^\prime)^2\right\}-2a\gamma^2\gamma^\prime+\gamma^3
\label{T3}\\
T_4&:=&8a^3\gamma\left\{\gamma^2\gamma^{\prime\prime\prime}
+4\gamma\gamma^\prime\gamma^{\prime\prime}+(\gamma^\prime)^3\right\}
+4a^2\gamma^2\left\{2\gamma\gamma^{\prime\prime}
+(\gamma^\prime)^2\right\}+4a\gamma^3\gamma^\prime-\gamma^4
\label{T4}
\end{eqnarray}
where the first two terms on the r.h.s.\ of~(\ref{T34}) parallel
those in~(\ref{conv}) and the remainder are grouped in $T_3$,
which is cubic in $\gamma$, and $T_4$, which is quartic.

At first sight, it was hard to spot the pattern of higher-order terms
in~(\ref{T3},\ref{T4}). It emerged when we discovered that
\begin{equation}
T_3=\gamma\left(2a{d\over d a}-1\right)T_2\,;\quad
T_2:=a^2{d\over dx}{\gamma^2\over a}
=\gamma\left(2a{d\over d a}-1\right)\gamma\,.
\label{found3}
\end{equation}
It was then natural to suppose that
\begin{equation}
T_4=\gamma\left(2a{d\over d a}-1\right)T_3
\label{found4}
\end{equation}
and to verify that this indeed gives~(\ref{T4}). Hence we achieve
a highly systematic iteration
\begin{verbatim}
procedure DysonScalar(n); for m:=1:n do
<<for j:=1:3 do g(j+1,m):=for k:=j:m-1 sum g(m-k)*(2k-1)*g(j,k);
write G m:=11g(2,m)/6-6g(3,m)+6g(4,m)+if m=1 then 1; g(1,m):=6G m>>;
\end{verbatim}
which generates the 500-loop integer of Table~1 in merely 7
minutes. We will derive this procedure from the Hopf algebra
structure of the perturbation series $X$, but first we exhibit a
further consequence of this iteration.

\section{Momentum dependence {\sl via} the anomalous dimension}
\subsection{Propagator--coupling duality}
Finally we arrive at a fine duality, between the dependence of the
propagator
\begin{equation}
\Pi(q^2):={1\over q^2-q^2\Sigma(q^2)}
\label{Pi}
\end{equation}
on the logarithm $L:=\log(q^2/\mu^2)$, and the
dependence of the (suitably normalized) coupling constant
$\sqrt{a}:=g/(4\pi)^{d/4}$
on the dual of $L$, say $\lambda$, which we shall now identify.

First consider the Dyson-Schwinger equation
for $\Pi(q^2)$, which may be written
\begin{equation}
P\left(q^2{d\over d q^2}\right){1\over\Pi(q^2)q^2\sqrt{a}}
=-\Pi(q^2)q^2\sqrt{a}\,.
\label{Pq}
\end{equation}
Now consider the apparently baroque third-order equation~(\ref{3o})
for the anomalous dimension. Its origin is far simpler:
\begin{equation}
P\left(-2a\gamma(a){d\over d a}\right){1\over\sqrt{a}}=-\sqrt{a}
\label{Pg}
\end{equation}
with the {\em same\/} polynomial, of a different operator.
Thus the duality is as follows.
\begin{quotation}\noindent
{\bf Propagator-coupling duality} At fixed coupling $a_0$ let
\begin{equation}
\Pi(q^2)q^2\sqrt{a_0}=S(L);\quad L:=\log(q^2/\mu^2)\,.
\label{L}
\end{equation}
Let $\gamma(a)$ be the anomalous dimension at arbitrary coupling $a$.
Then
\begin{equation}
\sqrt{a}=S(\lambda)\,;
\quad \lambda:=\int_a^{a_0}{db\over2b\gamma(b)}\,.
\label{lambda}
\end{equation}
\end{quotation}

Thus we expose the physical reasons for the enormous speedups
that we have gained. The triple expansion in $a$, $L$ and $\varepsilon$,
employed in~\cite{exp,BK}, is obviously reducible to a double expansion
in $a$ and $L$, since we dispense with
dimensional regularization, by taking derivatives of the
Dyson-Schwinger equation. What was not so obvious, at the outset, is that
the double expansion in $a$ and $L$, underlying the procedure
{\tt HopfScalar} of sect.~2.2, is also unnecessary.
Now, thanks to the amazingly compact procedure~{\tt DysonScalar} of sect.~2.5,
we have a procedure for the perturbation series of $\gamma(a)$ by
iterative solution of the momentum-independent equation~(\ref{Pg}).

We had a strong intuition of the possibility of this second speedup,
by a huge factor of $O(n^2)$ at $n$ loops, arguing as follows.
Suppose that we knew the momentum dependence of the self energy, $\Sigma(q^2)$.
Then the anomalous dimension would be known, from
the derivative of the self energy at $q^2=\mu^2$, where $\Sigma(\mu^2)=0$.
Conversely, one expects that a knowledge of the dependence of $\gamma(a)$
on the coupling $a$ is sufficient to determine the momentum dependence of
the self energy, and hence that it is unnecessary to expand in both
$a$ and $\log(q^2/\mu^2)$.
The dictionary~(\ref{L},\ref{lambda}) establishes this fact.

In summary: the self energy must be renormalized; the anomalous
dimension encodes the infinities that have been subtracted. From
the momentum dependence of the self energy one can find the
anomalous dimension; in massless theories the converse is true.
Note that we can read the duality as saying $$  \log(q^2/\mu^2)
=\int_{ q^2\Pi(q^2) }^1 \frac{dx}{x} \frac{1}{\gamma(x^2 a)} $$ or
as\\ {\bf Theorem 1}: $$\frac{d\log(1-\Sigma)}{ d\log
q^2}=\gamma(a/(1-\Sigma)^2).$$ We now derive this result
rigorously.
\subsection{Proof of the duality}
We have to prove Theorem 1. We will rely on two propositions. The
first exhibits how the Hopf and Lie algebra structure of
perturbation theory interferes with the equation of motion --the
Dyson--Schwinger equation--  which is the defining equation
$X=aB_+(1/(1-X))$ for the Hopf algebra element $X$. The second
proposition establishes the recursion {\tt DysonScalar}.

 Let
$X=\sum_{k=1}^\infty a^k X_k$ be a formal series in the Hopf
algebra $H$ of undecorated rooted trees , such that the counit
$\bar{e}$ vanishes, $\bar{e}(X)=0$, with $X_k\in H$ and $a$ be the
coupling constant, and as announced above let $X$ be defined by $$
X=a B_+(1/(1-X)),$$ where we write $1/(1-X)=\sum_{j=0}^\infty
X^j$. This determines the $X_k\in H$ uniquely.  The first four
$X_k$ read \begin{eqnarray*} X_1 & = & B_+(e)\\ X_2 & = &
B_+(X_1)\\ X_3 & = & B_+(X_2)+B_+(X_1X_1)\\ X_4 & = &
B_+(X_3)+2B_+(X_1X_2)+B_+(X_1X_1X_1),
\end{eqnarray*}
which are linear combinations of rooted trees, and the unit $e$ of
the algebra represents the empty tree.

Note that the weights in $X_4$ are not the Connes-Moscovici
weights, but field--theoretic weights.

Let $F_T(U)$ be the befooting operator (with $F_{u_1}$ the usual
befooting operator of \cite{Chen,Coho}) which sums over all ways
of removing the tree $T$ from $U$, extended by linearity: for
$Y=cT_1+dT_2$ a linear combination of rooted trees (ie an
arbitrary element of the linear basis of $H$), let us set $$
F_Y(U)=cF_{T_1}(U)+dF_{T_2}(U).$$\\ {\bf Prop.1}: The above
coefficients $X_i$ of the series $X$ fulfill
$$F_{X_m}(X_k)=[2(k-m)-1)]X_{k-m},\;k>m\geq 1.$$\\ Proof: The
structure of the series $X$ gives the following form to the
coefficients $X_k$:
\begin{equation}
X_{k+1}=B_+\left[\sum_{r=1}^k\sum_{{s_1i_1+\ldots+s_ri_r=k\atop
0<i_1<\ldots<i_r}}X_{i_1}^{s_1}\ldots
X_{i_r}^{s_r}\frac{r!}{s_1!\ldots s_r!} \right],\;\forall k\geq 1
.\label{xk1}\end{equation}

Note that by
definition $F_{X_m}(X_k)=0$, $m\geq k$ and that $F_{X_m}$ is a
derivation, $F_{X_m}(UV)=F_{X_m}(U)V+UF_{X_m}(V)$. Also,
\begin{equation}F_{X_m}(X_{m+1})=X_1,\;\forall m\geq 1,\label{start}\end{equation} as
\begin{eqnarray*}
F_{X_k}(X_{k+1}) & = & F_{X_k}\left(
B_+\left[\sum_{r=1}^k\sum_{{s_1i_1+\ldots+s_ri_r=k\atop
0<i_1<\ldots<i_r}}X_{i_1}^{s_1}\ldots
X_{i_r}^{s_r}\frac{r!}{s_1!\ldots s_r!} \right] \right)\\
 & = & F_{X_k}(B_+(X_k))=X_1.\end{eqnarray*} For any $m$ in $F_{X_m}(X_i)$, we assume the
proposition holds for all $X_i$, $i\leq k$. We show that it holds
for $X_{k+1}$, $\forall m$. For any $m$, Eq.(\ref{start}) provides
the start of the induction.

Now, $F_{X_m}$ and $B_+$ have the commutator
$$[F_{X_m},B_+](X_k)=\delta_{m,k}B_+(e),$$ by the definition of
$F_{X_m}$ ($F_{X_m}$ removes feet, $B_+$ grows at roots, only at
$m=k$ this interferes in the indicated manner), which generalizes
to any product $\prod_i X_{r_i}$ of elements $X_{r_i}$ so that the
interchange of the befooting operator generates some derivatives
of this product: $$F_{X_m}B_+\left[\prod_i X_{r_i}\right]
=B_+\left[F_{X_m}(\prod_i X_{r_i} )+\sum_j
\delta_{m,r_j}\prod_i^{\hat{r_j}} X_{r_i}\right], $$ where the
superscript $\hat{r_j}$ at the product
 indicates an omission of the factor $X_{r_j}$.

We can now interchange the action of the $B_+$ and the $F_{X_m}$
operator in the expression for $F_{X_m}(X_{k+1})$, and use the
assumption to find the desired result. Let us work out an example
before we give some more details. $$
X_6=B_+[X_5+2X_1X_4+2X_2X_3+3X_1^2X_3+3X_1X_2^2+4X_1^3X_2+X_1^5].$$
For $F_{X_m}(X_6)$ we thus find (using $F_{X_m}(X_1)=0, \forall
m$) \begin{eqnarray*}
 & & B_+[F_{X_m}(X_5)+2X_1F_{X_m}(X_4)+2F_{X_m}(X_2)X_3+2X_2F_{X_m}(X_3)+
3X_1^2F_{X_m}(X_3)\\ & & +6X_1X_2F_{X_m}(X_2)+4X_1^3F_{X_m}(X_2)]
+ \;\mbox{commutator terms}.\end{eqnarray*} To be specific, let us
set $m=2$ to get, using the assumption of the induction, $$
B_+[5X_3+6X_1X_2+2X_2X_1+ 3X_1^2X_1],$$ while the commutator terms
deliver $$B_+[2X_3+6X_1X_2+4X_1^3].$$ Altogether, we get $$
B_+[7X_3+14X_1X_2+7X_1^3]=7B_+[X_3+2X_1X_2+X_1^3]=7X_4,$$ as
desired.

Let us now turn to the general case. Let us consider a monomial of
degree $r$ contributing in the sum in Eq.(\ref{xk1}). It suffices
to consider the generic case $s_1=\ldots=s_r=1$, if there are
higher powers $s_i>1$, the fact that $F_{X_m}$ is a derivative
ensures that the same argument holds and that the extra $s_i!$
factors in the denominator will be appropriately cancelled.

In the generic case the coefficient of the monomial is $r!$. Let
us first consider the case $r=2$. Contributions to $X_aX_b$ say,
with $a+b=k-m$, are coming from
$$2!F_{X_m}(B_+(X_{a+m}X_b+X_aX_{b+m})).$$ By assumption of the
induction, this delivers, for the non-derivative terms,
$$2![2(k-m)-2]B_+(X_aX_b),$$ while from the derivative terms, we
get $$3!B_+(X_aX_b),$$ which adds up to
$$2![2((k+1)-m)-1]B_+(X_aX_b),$$ as desired. In general, for
arbitrary $r$, the simple fact that $r!(r+1)=(r+1)!$ ensures that
each term in the sum homogenously factors $[2((k+1)-m)-1]$, and
summing over all monomials and applying the $B_+$ operator, we
get, by Eq.(\ref{xk1}) backwards,
$F_{X_m}(X_{k+1})=[2((k+1)-m)-1]X_{k+1-m}$. $\Box$\\[5mm] Now, let
$\sigma^n_m:=\lim_{\epsilon\to 0}\epsilon^n\left[\phi\circ
S\star[Y^n]\right](X_m)$, with antipode $S$ and grading $Y$ and
convolution product $\star$ as in \cite{BK}.\\ Then, the procedure
{\tt DysonScalar} is\\ {\bf Prop.2}:
$$\sigma^n_m=\sum_{i=1}^{m-1}[2(m-i)-1]\sigma_i^1\sigma_{m-i}^{n-1}.$$
Here, $\phi$ is the unrenormalized character on the Hopf algebra
$\phi={\bf\Sigma}\mid_{q^2=1}$. With its help and the Hopf algebra
automorphism $\Theta_\rho:H\to H$ of \cite{RHII},
$$\Theta_\rho(h)=e^{\rho {\rm deg}(h)}h,\;\forall \rho\in H$$
(${\rm deg}$ is the degree: ${\rm deg}(X_m)=m$) we can write the
renormalized character: the character ${\bf\Sigma}$ is given by $$
{\bf\Sigma}(q^2/\mu^2)=\phi\circ \left[S\star
\Theta_{-\epsilon\log(q^2/\mu^2)}\right ].$$ The Taylor expansion
in $\log(q^2/\mu^2)$ is then  solely a study of the expansion of
the Hopf algebra automorphism $S\star
\Theta_{-\epsilon\log(q^2/\mu^2)}$ which can be written, setting
$z=\log(q^2/\mu^2)$, as $$ S\star
\Theta_{-\epsilon\log(q^2/\mu^2)}=\sum_{i=1}^\infty \frac{S\star
[Y^i]}{i!}(-\epsilon z)^i.$$

Also, we note that quite generally
\begin{equation}
\epsilon^{n}\phi(S\star Y^m[X_j])={\cal O}(\epsilon^{n-m}),\;
\forall X_j,\label{asy}
\end{equation}
which follows from the fact that ${\bf\Sigma}$ exists in the limit
$\epsilon\to 0$.

Proof: We use Prop.1 and induction on $m$. We only consider the
case $n=2$, the other cases follow identically. The start of the
induction involves an explicit check of properties of $\phi(X_1)$,
$\phi(X_2)$, which can be done easily using $\phi(X_1)=B_1,
\phi(X_2)=B_1B_2,$ where $$ B_i=\int d^Dk
\frac{1}{[k^2]^{1+i\epsilon}(k+q)^2}\;{|}_{q^2=1}. $$ We
concentrate on the induction step. Assume $$ \epsilon^2
\phi[S\star Y^2[X_j]]=\epsilon^2\sum_{i=1}^{j-1}\phi[S\star
Y[X_1]]\phi[S\star Y[F_{X_i}(X_j)]]$$ for all $j\leq m$. We want
to prove it for $m+1$. Again, we use the presentation
Eq.(\ref{xk1}) for $X_{m+1}$, for which we write in shorthand
$X_{m+1}=B_+(X_m+\sum\prod_i X_i)$, empasizing a decomposition
into product terms and the single term $X_m$ which is linear in
generators in the sum involved in this equation.

We then have
\begin{eqnarray}
\epsilon^2\phi(S\star Y^2[X_{m+1}]) & = & \epsilon^2\phi[S\star
Y^2[B_+(X_m+\sum\prod_i X_i)]]\nonumber\\
 & = & \epsilon^2\phi(S\star Y^2[B_+(X_m)])+\epsilon^2\widetilde{B_+}\phi(
S\star Y^2[\sum\prod_i X_i]).\label{new}
\end{eqnarray}
Here, $\widetilde{B_+}(\phi)\equiv \phi\circ B_+$ and we use $$
S\star Y^2\circ B_+=m\circ(S\otimes Y^2)\circ(id\otimes
B_+)\circ\Delta $$ (as $Y^2(e)=0$ and where $m$ is the product,
$\Delta$ the coproduct) which further can be written as $$
m\circ(id\otimes B_+)\circ(S\otimes Y^2)\circ\Delta, $$ due to the
fact that for the commutator between $Y,B_+$ we have
$[Y,B_+]=B_+$. Hence, interchanging $B_+$ with $Y^2$ produces only
commutator terms involving $\epsilon^2\phi\circ S\star Y^i$, with
$i=0,1$, which vanish by Eq.(\ref{asy}) in the limit $\epsilon\to
0$ .

Let us treat the two terms on the rhs in the last line of
Eq.(\ref{new}) separately. For the second term, we use that
$S\star Y^k[\prod_i X_{i_j}]=0$ if the product has  more than $k$
factors. To see this, consider, for example (using Sweedler's
notation), \begin{eqnarray}S\star Y[X_iX_j] & = & S[X_i^\prime
X_j^\prime]X_i^{\prime\prime} X_j^{\prime\prime}[{\rm
deg}(X_i^{\prime\prime})+{\rm
  deg}(X_j^{\prime\prime})]\nonumber\\
& = & S\star id(X_i)S\star Y(X_j)+S\star Y(X_i)S\star
 id(X_j)=0,\label{van}
\end{eqnarray} and similarly for higher powers of $Y$. Hence, $S\star Y^2$
vanishes on products of more than two arguments, and on two
arguments we have, $S\star Y^2(X_iX_j)=2S\star Y(X_i)S\star
Y(X_j)$. (In general, $S\star Y^n$ behaves like a $n$-th
derivative acting on a product of functions which vanish at the
origin, for example $$(fg)^{\prime\prime}=f^{\prime\prime}g+
2f^{\prime}g^{\prime}+fg^{\prime\prime} =2f^{\prime}
g^{\prime}.)$$ Hence, the second term on the rhs above delivers $$
2\epsilon^2\sum_{i+j=m}\widetilde{B_+}\phi[S\star
Y[X_i]]\phi[S\star Y[X_j]]. $$ For the first term on the rhs we
find $$ \epsilon^2 [\phi\otimes\phi](S\otimes Y^2)(id\otimes
B_+)\Delta(X_m)  =m[(id\otimes
\widetilde{B_+})(\phi\otimes\phi)(S\otimes Y^2)\Delta(X_m)]. $$
Here we interchanged $B_+$ with the $Y$ operator as before. The
assumption of the induction  is $$ (\phi\otimes\phi)\circ(S\star
Y^2)\circ \Delta(X_j)= (\phi\circ S\star Y\otimes\phi\circ S\star
Y)\circ\Delta(X_j). $$  From this assumption  and from Prop.1 we
can write for the above expression $$ \sum_i
[2(m-i)-1]\left[\phi\circ S\star
Y(X_i)\right]\,\,\left[\tilde{B_+}(\phi)\circ S\star Y
(X_{m-i})\right]. $$ Adding the results of the first and second
terms  for the rhs of (\ref{new}), we find $$
\sum_i[2((m+1)-i)-1]\left[\phi\circ S\star
Y(X_i)\right]\,\,\left[\tilde{B_+}(\phi)\circ S\star
Y(X_{m-i})\right]. $$ Using that $S\star Y$ vanishes on products
by Eq.(\ref{van}), we can combine the resulting expressions to the
desired result, by adding zero in form of adding all the product
terms in Eq.(\ref{xk1}) for $X_{m+1-i}$ to eliminate the $B_+$
operator in favour of the increased degree $m+1-i$. This proves
the proposition.$\Box$\\

\noindent With these tools, the proof of Theorem 1 becomes a mere
comparison of coefficients of two series in an infinite number  of
variables. Indeed, in both series we can express any appearance of
$\sigma^n_m$ by Prop.~2 as a product $\Pi_{i=1}^n \sigma^1_{m_i}$,
$\sum_i m_i=m$. Hence, Theorem 1 provides two power series in an
infinite set of variables $\sigma^1_i$. Note that a monomial in
degree $m$ in these variables is always accompanied by the factor
$z^{m-1}$.

Let us set
$$\sigma:=\sum_{k=1}^\infty\sum_{n=1}^\infty\frac{\sigma_k^n
z^n}{n!},$$ and $$\sigma^{(i)}:=\sum_{k=1}^\infty \sigma^i_k.$$
Then, Theorem 1 amounts to showing that \begin{equation}
\sum_{k=1}^\infty\sum_{n=0}^\infty
\frac{\sigma_k^{n+1}z^n}{n!}\frac{1}{1-\sigma}= \sum_{m=1}^\infty
\sigma_m^1\frac{1}{(1-\sigma)^{2m}}.\label{prove}\end{equation}

Using Prop.2 and elementary algebra one finds the following
coefficients at orders $z^0,z^1,z^2/2,\ldots$ for the lhs:
\begin{eqnarray*}
z^0: & & \sigma^{(1)},\\ z^1: & &
\sigma^{(1)}\sum_{r_1=1}^\infty(2r_1)\sigma^1_{r_1},\\
\frac{z^2}{2}: & & \sigma^{(1)}\sum_{r_1,r_2=1}^\infty
(2r_2)(2r_1+2r_2) \sigma^1_{r_1}\sigma^1_{r_2},\\ \cdots & &
\cdots,
\end{eqnarray*}
which for example implies that the coefficient of the term
$[\sigma_1^1]^n z^{n-1}/(n-1)!$ is simply $2^{n-1}(n-1)!$. The
crucial step in all this is to recognize how the terms coming from
the expansion of $1/(1-\sigma)$ cancel the `$-1$'-term coming from
Prop.2, ie the term $-\sum_i \sigma^1_i\sigma^{n-1}_{m-i}$ in $$
\sigma_m^n=\sum_i[2(m-i)-1]\sigma^1_i\sigma_{m-i}^{n-1}.$$ A
similar phenomenon happens in the Taylor expansion of $$
\frac{1}{(1-\sum_{i=1}^k \frac{\sigma^{(i)} z^i}{i!})^{2m}}$$ to
the $k$-th order, involved in the rhs of (\ref{prove}). For
example, at order $z^2$, $k=2$, the `$-1$'-term in $(2m-1)$
generated by the second derivative does the same job, and in
general the lhs and the rhs of (\ref{prove}) give the same series
in $z$. $\Box$\\

\section{Conclusions}
Returning to the questions in the introduction, we answer them as
follows:

\noindent 1.~The non-perturbative problems whose expansions were
studied in \cite{exp,BK} are defined by (14,18). For the Yukawa
problem, a solution in closed form is given by (1), for the
anomalous dimension. The duality of Theorem 1 then gives the
propagator as a limit of integration of an integral that must
evaluate to $\log(q^2/\mu^2)$. The particular solution of this
duality is given by (37,38), in terms of the complementary error
function. The same form of duality applies in the scalar case,
though for this we have not yet obtained an explicit solution to
the 4th-order equation that results from (53--55). We have,
however, found a way of developing the perturbation series almost
as efficiently as in the Yukawa case, attaining 500 loops in a few
minutes.

\noindent 2.~The obstacle in (5) at $\tilde\gamma=-2$, i.e.~at
coupling $g=2(2\pi)^{3/2}\approx31.5$ in the Yukawa case, was
surmounted, thanks to the parametric non-perturbative solution
(1). In the scalar case, the obstacle in (48) at $\gamma=-1$ is
surmountable, by numerical solution of the third-order equation
for the anomalous dimension. In the Yukawa case, we are able to
compute strong-coupling results to very high accuracy and use them
as stringent tests of the Borel resummation in \cite{Jen}; in the
scalar case, we have confidence in such strong-coupling
resummations, but cannot yet surpass them with the ease afforded
by (1,37,38).

Let us now finish this paper by a discussion of the difficulties
one confronts in general. Our starting Dyson--Schwinger equations
were addressing Feynman graphs representing only the Hopf algebra
of undecorated rooted trees. To treat the general case, the first
observation is that there would be various operators $B_+$,
indexed by labels corresponding to primitive graphs without
subdivergences, of the form $$ \widetilde{B_+}^x=\frac{a^{{\rm
deg}(x)}}{\pi^{2{\rm deg}(x)}}\int d^4l\frac{F^x(l^2,l\cdot q,
q^2)}{l^2},$$ where ${\rm deg}$ gives the loop number of the
decoration $x$, $F^x$ is some integral kernel obtained from the
skeleton graph $x$, and the corresponding integral operator
contains a sum $\widetilde{B_+}=\sum_x \widetilde{B_+}^x$. Here,
the previous representation of the $\widetilde{B_+}$ operator as
$$ \frac{a}{\pi^2}\int d^4l \frac{q\cdot l}{l^2(l+q)^2}$$ would be
just the first term in such a series.

A further complication is given by the fact that allowing for
internal self-energies and vertex-corrections one would have to
consider a coupled system of Dyson--Schwinger equations for those
functions. Still, all this is conceptually not too drastic a
challenge, and Hopf and Lie algebra structures for the full theory
are readily available \cite{DK,more}.

But the real difficulty resides in the fact that these new
$\widetilde{B_+}^x$ operators offer a choice of as many places as
there are internal lines and vertices for the self-energies and
vertex corrections to be inserted into $x$. This is in sharp
contrast to the Dyson--Schwinger equation studied here which was
carefully designed so that it was always the same propagator into
which the self-energy was inserted (see Figure 1).

It would be very convenient if the characters of the full Hopf
algebra of Feynman graphs would turn out to be independent of such
choices, alas, in dimensional regularization, they are not.
Nevertheless, they fail to be so in a most interesting way, as
mentioned in \cite{review}: only lower order pole terms are
affected by such choices, and a proper understanding of the operad
structure of Feynman graphs \cite{review} should ultimately allow
to generalize the methods presented here to the full theory.

\raggedright

\newpage

{\bf Table 1}\quad Coefficient of the scalar perturbation series with 1675
digits at 500 loops
{\footnotesize\begin{verbatim}
G500=206261451966080541451119356265266407905816117576895601520616328670543304097
62369668214104674763068056454522518617422020409397336434904863988900797769773644
47129884863324773181376863120291798830884688213932683869821267125662274428136514
68974978228592824043044373847281757207937081063432528806815509319762088807291996
54549245884853496719417048678199825379018355919198123075612308008976364608893906
00835837012056033720017238115336850340799075684336975651857656078799282745256216
85768456030809283727097722850488278232311177219444745322287340871435443707536590
64304859950724683157717734493071321199539578218428617617722892100276682781401203
04983974209704793621909710059353724523231635766062166284812903992269403282699432
81718327508638643305481989940132234093616573076862094588977827344981584305605437
66475002382217933275761312682929603923397580260987048907414858143897114762331252
08694985337972553885925402003826420205441859988844001867088083850782378303677991
14077650584544145709672328391394562704209221732180879565868213522109303655045186
92714017665002971967455255310508358729281544729403249398746232320441525286283859
23093041626365262630100048817481274793707664791767175677240144896307853488347045
21622394885797995125083750860330519417878429051836575477220881369445751634601965
33191009573619480068718080810533581305996863996579338522874547127421808710757882
86996199556804886954946559116947132125235605586627322129268965041445488085748194
82341875039156647569797757032552836429751077302524927736861138479038542006096835
73747720303607608007740173613335602076396299832459826245418033598839559699294537
37336134624690115674194793212055897162647586497730033948880084738561472545509216

\end{verbatim}}

\end{document}